# Spontaneous sarcomere dynamics

Stefan Günther[1, a)] and Karsten Kruse[1, b)]
*Theoretical Physics, Saarland University, 66041 Saarbrücken, Germany*

(Dated: 24 November 2010)

Sarcomeres are the basic force generating units of striated muscles and consist of an interdigitating arrangement of actin and myosin filaments. While muscle contraction is usually triggered by neural signals, which eventually set myosin motors into motion, isolated sarcomeres can oscillate spontaneously between a contracted and a relaxed state. We analyze a model for sarcomere dynamics, which is based on a force dependent detachment rate of myosin from actin. Our numerical bifurcation analysis of the spontaneous sarcomere dynamics reveals notably Hopf bifurcations, canard explosions, and gluing bifurcations. We discuss possible implications for experiments.

In skeletal muscles, the elementary force producing units are sarcomeres. They consist of interdigitating filaments of molecular motors and actin filaments. Upon activation by a neural signal, the motors move along the actin filaments and thus shorten the sarcomere. Elastic elements maintain the regular arrangement of motor and actin filaments. It has been found experimentally that in the absence of neural signals, sarcomeres can spontaneously oscillate in length. We study a possible mechanism of spontaneous sarcomere oscillations, which is based on a dynamic instability of motors coupled to an elastic element. We find a large variety of nonlinear behavior ranging from simple oscillations to excitable and chaotic dynamics. Our results indicate possible ways to experimentally test the mechanism we propose for spontaneous sarcomere oscillations.

## I. INTRODUCTION

Striated muscles consist of fibers, called myofibrils, which are formed by a periodic arrangement of interdigitating myosin filaments and actin filaments. Myosin is a motor protein, which can convert the chemical energy released during the hydrolysis of adenosine-tri-phosphate (ATP) into mechanical work when it interacts with actin filaments. Importantly, these filaments have two structurally distinct ends and in this way determine the direction of the force generated by myosin motors. A unit cell of the periodic myofibril structure is called a sarcomere and presents the basic contractile unit of striated muscles, see Fig. 1A. In a sarcomere, the interaction of the myosin motors with the polar actin filaments leads to unidirectional sliding between the two kinds of filaments and induces sarcomere and hence muscle contraction. The structural integrity of sarcomeres is assured by elastic elements, notably Z-discs, M-lines, and titin molecules.

In an organism, striated muscles usually contract as a result of an increased concentration of $Ca^{2+}$ ions subsequent to a neural signal. However, for example, the muscles responsible for beating the wings of some flies contract at a larger frequency than neurons are capable of firing action potentials[2]. Hence, other than neural mechanism must generate the necessary periodic muscle contraction. Remarkably, it has been found that sarcomeres can oscillate spontaneously, even in the absence of $Ca^{2+}$[3,4].

Possible mechanisms for spontaneous mechanical oscillations are built on dynamic instabilities of molecular motors[5–10]. How ensembles of molecular motors coupled to an elastic element can spontaneously oscillate is most easily seen in the case of force-dependent detachment rates. Consider motors, which attach to a polar filament. Assume that the motors compress an elastic element as they move along the filament. Consequently, the load on the motors increases. If the rate of motor detachment from the filament grows with the applied force, the detachment of one motor can induce a detachment avalanche. Indeed, with the detachment of each motor, the load on the remaining attached motors instantly increases accelerating their detachment even further. After all the motors have detached, the elastic element relaxes and the cycle starts again. Spontaneous mechanical oscillations caused by molecular motors have been invoked to explain various physiological processes, for example, the beating of cilia and flagella[5,11], the rocking of mitotic spindles during asymmetric cell division[12], and chromosome oscillations during cell division[13,14].

We employed motor induced dynamic instabilities to investigate spontaneous sarcomere oscillations and showed that such a mechanism can generate relaxation waves similar to those observed in myofibrils[15]. In the

---

[a)] stefan.guenther@physik.uni-saarland.de;
http://www.uni-saarland.de/~stefan;
Present address: EMBL Heidelberg
[b)] k.kruse@physik.uni-saarland.de;
http://www.uni-saarland.de/fak7/kruse

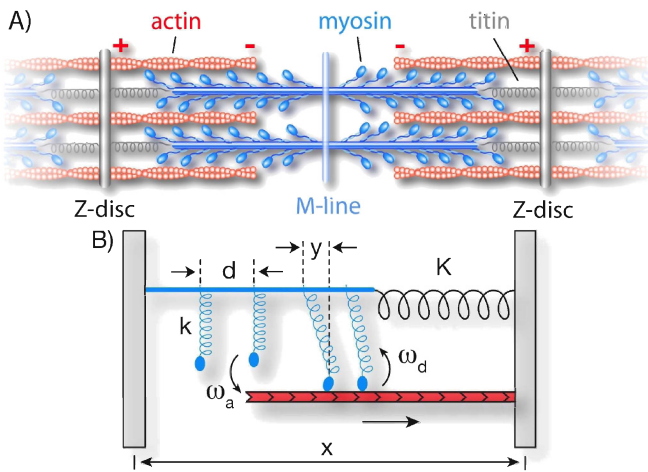

FIG. 1. A) Illustration of a sarcomere. Bipolar myosin filaments interdigitate with actin filaments, which are attached with their plus-ends to Z-discs. Upon activation of the motors, the actin filaments are pulled towards the M-line resulting in sarcomere contraction. B) Illustration of the model describing the dynamics of a half-sarcomere. The parallel actin and myosin filaments are, respectively, replaced by single effective filaments. Motors are attached via elastic springs of stiffness $k$ with extension $y$ to the common backbone, separated from each other by a distance $d$. These effective motors are processive and have a well-defined force-velocity relation, see text. The motors' detachment rate is force-dependent. A spring of stiffness $K$ accounts for the elastic components of the structure.

present work, we perform a numerical bifurcation analysis of the spontaneous dynamics of sarcomeres. We start by analysing the dynamics of a half-sarcomere and find that a supercritical Hopf bifurcation is followed by a canard explosion and a subsequent secondary saddle-node bifurcation resulting in a logarithmic divergence of the oscillation period. Sarcomeres show phase-shifted oscillations of the forming half-sarcomeres and can present global (gluing) bifurcations. The phase-shifted oscillations are at the basis of the traveling relaxation waves in myofibrils and can be used to generate a simple self-organized swimmer[16].

## II. DYNAMICS OF A HALF-SARCOMERE

### A. Dynamic equations

We will start our analysis of the dynamics of sarcomeres by considering one half of a sarcomere. While it is physiologically not possible to cut a sarcomere along the M-line into two functional halves, from a dynamic point of view, they are the basic units of a myofibril. We describe a half-sarcomere as consisting of effective motor heads, which are elastically attached to a common backbone, see Fig. 1B. The linker springs have a stiffness $k$ and neighboring springs on the backbone are separated by a distance $d$. Each motor head $i$ can attach to a polar filament. When a motor is attached, it moves along the filament subject to a force-velocity relation $v(f) = v_0(1 - f/f_s)$, where the force $f = ky_i$ is determined by the linker extension $y_i$, $v_0$ is the velocity of a free motor, and $f_s$ is the force at which the motor stalls. Motors in the overlap region of the motor and the polar filament attach to the polar filament at a rate $\omega_a$ and detach at a force-dependent rate $\omega_d(y_i)$. The elastic elements of a half-sarcomere of length $x$ are accounted for by a spring of stiffness $K$ with a rest length $\ell_0$. We will now briefly introduce the equations governing the dynamics of this system. The equations are based on a mean-field analysis and a detailed derivation can be found in Ref.[15].

The effective motors mentioned above result from averaging over the $M$ myosin motors in a slice of width $d$ perpendicular to the sarcomere's long axis. The size of $d$ is determined by the distance between adjacent myosin binding sites on an actin filament. In the mean-field approximation, we assume that all linker springs have the same extension, $y_i = y$ for all $i$. Together, the motors generate thus a force $f_m = -N(x)Qky$, where $N(x) = (\ell_p + \ell_m - x)/d$ is the number of motors in the overlap region of the motor filament and the polar filament, which are of length $\ell_m$ and $\ell_p$, respectively, and $Q$ is the probability of a motor to be bound. Force balance now gives a first dynamic equation. In the absence of external forces, the motor force $f_m$, the elastic force $f_e = -K(x-\ell_0)$, and the effective friction force $f_f = -\xi \dot{x}$ sum up to zero, $f_m + f_e + f_f = 0$, which implies

$$\xi \dot{x} = -N(x)Qky - K(x - \ell_0) \quad . \qquad (1)$$

The second dynamic equation is based on the geometrical relation $\dot{y}_i = v(y_i) + \dot{x}$. In the mean-field approximation, $\dot{y}$ can be estimated by $y \cdot \omega_d(y)$[12], so that

$$\dot{x} = y \cdot \omega_d(y) - v(y) \quad . \qquad (2)$$

From Eqs. (1) and (2), $x$ can be eliminated by first equating the respective right hand sides. Solving the resulting equation for $x$, computing the derivative with respect to time, and using Eq. (2) then leads to

$$\dot{y} = \frac{g(y) \cdot [Qy - \kappa]^2/\kappa - \dot{Q}y \cdot [\Lambda - \zeta g(y)]}{\zeta g'(y) \cdot [Qy - \kappa] + Q \cdot [\Lambda - \zeta g(y)]} \quad . \qquad (3)$$

Here, $g(y) = y(\gamma \omega(y) + 1) - 1$ and we have scaled time by $\omega_a$ and space by $f_s/K$, yielding the dimensionless parameters $\gamma = f_s \omega_a/(kv_0)$, $\kappa = dK/f_s$, $\Lambda = \omega_a(\ell_0 - \ell_m - \ell_p)/v_0$ and $\zeta = \xi \omega_a/K$.

It remains to determine the time evolution of the probability $Q$ for a motor to be bound. An effective motor is attached to the polar filament, if at least one of the constituting motors is bound. Assuming an exponential dependence of the detachment rate on the force for an individual motor, the rate's force dependence of an effective motor can be expressed as

$$\omega(y) = \left[(\omega_0 \exp(-\lambda|y|) + 1)^M - 1\right]^{-1} \quad . \qquad (4)$$

Here, $\omega_0$ is the fraction of the attachment and detachment rate of an unloaded individual motor while $\lambda > 0$ determines the sensitivity of the detachment rate to the force acting on an effective motor, which is built from $M$ myosin motors. This force is equal to the force the motor exerts on the filament. Thus, $Q$ evolves according to[12]

$$\dot{Q} = (1 - Q) - Q \cdot \omega(y) \quad . \tag{5}$$

Stationary states $(Q_s, y_s)$ of the dynamic equations (3) and (5) are determined by either

$$g(y_s) = 0 \quad \text{and} \quad Q_s = (1 + \omega(y_s))^{-1} \tag{6}$$

or by

$$Q_s y_s = \kappa \quad \text{and} \quad Q_s = (1 + \omega(y_s))^{-1} . \tag{7}$$

Let us note, that the extension $x$ of the half-sarcomere is not stationary for $y_s$ and $Q_s$ given by Eqs. (7). In fact, Eqs. (1) and (2) imply in this case $\dot{x} = \text{const}$. We note further, as $\dot{x} < 0$, the element shortens. This shortening should obviously stop at a certain minimal half-sarcomere extension, but this process is not captured by the dynamic equations (3) and (5). We will now discuss the stability of the stationary states $(Q_s, y_s)$.

### B. Linear stability

Let us first focus on the stationary states with $g(y_s) = 0$ as determined by Eqs. (6). For all parameter values, these equations have at least one solution with $y_s > 0$. We will consider in the following the effects of variations of the parameter $\kappa$, but the stationary states' stabilities can also be affected by varying the other system parameters[15]. Our findings apply, *mutatis mutandis*, to changes of their values, too. Since for large enough values of $\kappa$, the stationary states given by Eqs. (6) are stable, it turns out to be convenient to discuss the effects encountered by decreasing $\kappa$.

There are two critical values $\kappa_{\text{cr}}^{\pm}$ at which the stationary state changes stability. They are given by

$$\kappa_{\text{cr}}^{\pm} = \frac{(\zeta + \zeta\omega(y_s) + 2) Q_s y_s + C}{2(\zeta + \zeta\omega(y_s) + 1)}$$
$$\pm \frac{\sqrt{([\zeta + \zeta\omega(y_s)] Q_s y_s + C)^2 + 4 Q_s y_s C}}{2(\zeta + \zeta\omega(y_s) + 1)} \tag{8}$$

with

$$C = \Lambda Q_s \left(\omega(y_s) + 1 - y_s \omega'(y_s)\right) / g'(y_s) \quad . \tag{9}$$

Since the absolute value of the term involving the square root is always smaller than the modulus of the other term, either both values are positive or negative. The latter case is irrelevant because $\kappa$ must be positive. At $\kappa = \kappa_{\text{cr}}^+ \equiv \kappa_h$, we find a backward Hopf bifurcation, see Fig. 2, while for $\kappa = \kappa_{\text{cr}}^-$ the system undergoes a forward Hopf bifurcation. For the parameter values used in Fig. 2 the bifurcations at $\kappa = \kappa_{\text{cr}}^+$ and $\kappa = \kappa_{\text{cr}}^-$ are supercritical and subcritical, respectively.

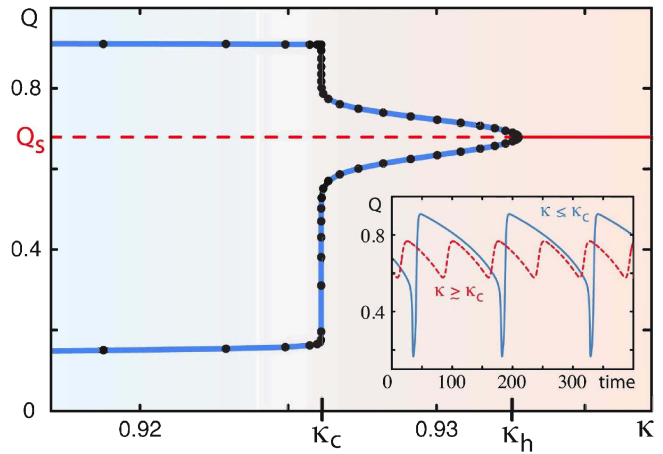

FIG. 2. Hopf bifurcation and canard explosion in a half-sarcomere. Red: stationary value $Q_s$ of the binding probability given by Eqs. (6). For $\kappa < \kappa_h = 0.933$, it is unstable and limit cycle oscillations emerge. Blue: extremal values of $Q$ during a cycle. At $\kappa \approx \kappa_c$ the oscillation amplitude explodes and relaxation oscillations are clearly detectable, see inset. Parameter values are $\gamma = 0.82$, $\Lambda = -82.4$, $\zeta = 6.6$, $M = 100$, $\omega_0 = 0.01$, $\lambda = 3$, and $\kappa = 0.926$ (inset, red dashed) as well as $\kappa = 0.9265$ (inset, blue). Numerical solutions are obtained using Auto07P[17] and the XPPAUT software[18].

Hopf bifurcations are possible under two conditions. On the one hand, it is necessary that $\kappa > Q_s y_s$, while on the other hand, we must have $y_s \omega'(y_s) > \omega(y_s) + 1$. The physical origin of the first inequality is the following: when the system is stretched by a distance $d$ while keeping the linker extension $y_s$ and the fraction of bound motors $Q_s$ constant, one effective motor leaves the overlap region of the polar filament and the motor filament. If the above condition is violated, then the decrease of the elastic force $f_e$ by $Kd$ is smaller than the corresponding reduction of the motor force $f_m$ and the elastic force will further extend the structure. If, on the contrary, the half-sarcomere is compressed by a distance $d$, then an analogous reasoning shows that the motors would win and further shorten the half-sarcomere. Incidentally, the stationary states given by Eqs. (7) with $Q_s y_s = \kappa$ define the marginal line separating states, respectively, fulfilling and violating $\kappa > Q_s y_s$. We will come back to this point below.

The second condition necessary for a Hopf bifurcation implies that the motors' detachment (or attachment) rates must depend on the force applied to the motor. This is the formal expression of the avalanche of motor detachments mentioned in Sec. I to be at the origin of the spontaneous oscillations.

Let us note, that at the bifurcation point, the critical

frequency $f_h$ is given by

$$f_h = \frac{1}{2\pi}\sqrt{\frac{\gamma\left(1+\omega(y_s)\right)\left(\kappa-Q_s y_s\right)^2}{\gamma\zeta\kappa\left(\kappa-Q_s y_s\right) - Q_s \Lambda \kappa/g'(y_s)}} \quad . \quad (10)$$

The frequency thus increases with the motors' detachment and attachment rates and increases with increasing stiffness of the spring $K$.

## C. Canard explosion of the limit cycle

Close to the bifurcation point, at $\kappa \approx \kappa_c < \kappa_h$, the limit cycle is strongly distorted and the amplitude explodes. For small changes in $\kappa$, $(\kappa - \kappa_c)/\kappa \lesssim 10^{-5}$, the amplitude in $Q$ increases three-fold, see Fig. 2. Also the oscillation shape changes significantly, see Fig. 2 inset. For $\kappa < \kappa_c$ the system clearly exhibits relaxation oscillations. Hence, the system contains two well-separated time scales. Typically, the dynamics of $Q$ is faster than that of $y$. However, the degrees of freedom corresponding to the two time-scales are really non-linear functions of $Q$ and $y$. Locally, the fast and slow degrees of freedom can be determined by diagonalizing the linearised dynamic equations. As a simple approximation, we use $Z = Q + y$ as the fast mode and $Q$ as the slow mode in the whole phase space.

The dramatic increase of the oscillation amplitude for small parameter changes is indicative of a canard explosion[19]. The reason behind this phenomenon is the system's excitability. Figure 3 shows the nullclines $\dot{Z} = 0$ in blue and $\dot{Q} = 0$ in red. The nullclines partition the phase space into regions with well-defined signs of $\dot{Z}$ and $\dot{Q}$. The $Q$-nullcline is independent of $\kappa$ and shows the typical S-shape common to oscillators presenting a canard explosion[20]. As long as $\kappa > \kappa_h$, the intersection point of the two nullclines, which corresponds to the stationary state $(Q_s, y_s)$, attracts the trajectories in its vicinity (not shown). For $\kappa_c < \kappa < \kappa_h$, small amplitude oscillations emerge, see Fig. 2 inset. However, in both cases, the system is excitable: for a large enough perturbation, the system first moves away from the respective fixed point or limit cycle and makes an excursion, before returning, see Fig. 3. For $\kappa \approx \kappa_c$, the excursions suddenly become part of the limit cycle leading to the explosion, see Fig. 2.

Let us note, that not all limit cycles in the system show a canard explosion. For example, in the limit $\Lambda \to 0$ and $\zeta \to 0$, the two values of $\kappa_h$ and $\kappa_{cr}^-$ approach each other and the limit cycles exist in an interval that is too small to allow for the explosion.

## D. Secondary bifurcations

We will now discuss the stationary states given by the solutions of Eqs. (7). We have mentioned already, that these states are not associated with a stationary half-sarcomere extension $x$; instead, the half-sarcomere's length shortens at constant speed. Similarly to our discussion of the stationary states determined by Eqs. (6) one might thus be inclined not to consider them further. However, as we will see, these states can have a profound effect on the limit cycles presented above.

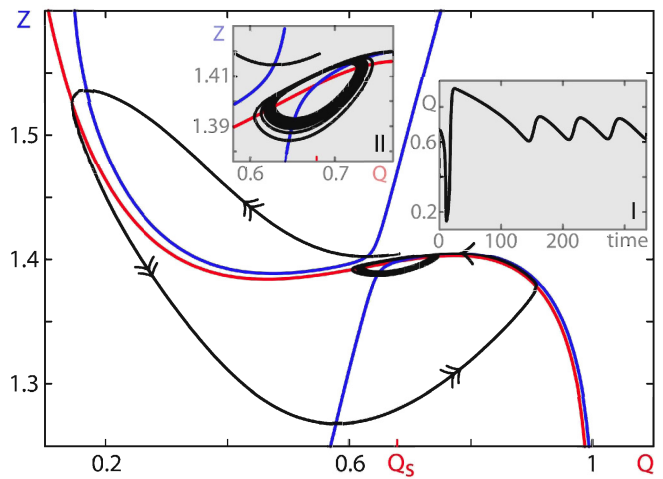

FIG. 3. An excited trajectory in phase space $(Z, Q)$ and nullclines for $Q$ (red) and $Z$ (blue) in the case of the half-sarcomere with $\kappa_c < \kappa = 0.93 < \kappa_h$. Inset I: Corresponding time course $Q(t)$. Inset II: Magnification of the phase space around the fixed point. Single arrows correspond to slow dynamics and double arrows correspond to fast dynamics. Other parameters are as in Fig. 2.

In contrast to the stationary states discussed so far, the solutions to Eqs. (7) exist only for values of $\kappa$ below a critical value $\kappa_{sn}$. At $\kappa = \kappa_{sn}$, two stationary states emerge through a saddle-node bifurcation. The value of $\kappa_{sn}$ can be determined by first eliminating $Q_s$ from Eqs. (7) giving $\kappa$ as a function of $y_s$ and then using $d\kappa/dy_s = 0$. This yields the following condition for $y_s^{sn} \equiv y_s(\kappa_{sn})$:

$$0 = y_s^{sn}\omega'(y_s^{sn}) - \omega(y_s^{sn}) - 1 \quad . \quad (11)$$

The sought for value is then $\kappa(y_s^{sn})$. In the limit $\exp(\lambda|y_s|) \gg \omega_0$, where $\omega(y_s) \approx [M\omega_0 \exp(-\lambda|y_s|)]^{-1}$, we obtain explicitly

$$y_s^{sn} = \lambda^{-1}\left[\mathcal{W}\left(M\omega_0/\exp(1)\right) + 1\right] \quad , \quad (12)$$

where $\mathcal{W}$ is the Lambert W-function.

A linear stability analysis shows that both states have a marginal direction with a growth exponent $\lambda_1 = 0$. The second growth exponent is $\lambda_2 = y_s\omega'(y_s) - \omega(y_s) - 1$, which has a different sign for the two states. Numerically, we find that the marginal direction can be non-linearly stable or unstable. Thus, the fixed points correspond to a saddle and a node, which might be stable or unstable. Remarkably, the stability of the node is intimately related to the saddle point given by Eqs. (6). Indeed, for $\kappa \to 0$ one of the two solutions of Eqs. (6) will have $Q_s \to 0$, while the other has $Q_s \to 1$, see Fig. 4. This implies that the stationary state of Eqs. (6) must, for some $\kappa$, be equivalent to the saddle or the node, because equal values of $Q_s$ imply equal values of $y_s$. Analyzing the flux lines of the dynamics for $Q$ we find that the crossing always occurs for the saddle. Furthermore, if the fixed point given by Eqs. (6) is stable for this value of $\kappa$, that is $\kappa_{\mathrm{sn}} > \kappa > \kappa_{\mathrm{h}}$, then the node is stable, otherwise it is unstable.

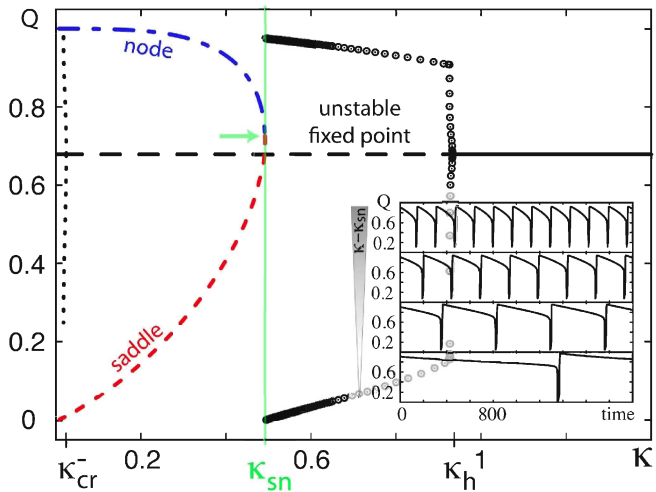

FIG. 4. Bifurcation diagram for a half-sarcomere. Stationary states and limit cycles of the half-sarcomere as a function of $\kappa$, respectively, represented by the stationary and extremal values of $Q$. The solutions of Eqs. (6) and the bifurcating limit cycles are represented by black, the solutions of Eqs. (7) by red and blue lines. The arrow indicates the saddle-node bifurcation point. At $\kappa_{\mathrm{cr}}^-$, we have a subcritical forward Hopf bifurcation. Inset: The oscillation period logarithmically diverges as $\kappa \to \kappa_{\mathrm{sn}}$. Parameters are as in Fig. 2.

As long as the saddle-node bifurcation occurs such that the limit cycle does not coexist with the saddle and the node for any value of $\kappa$, the "physically relevant" dynamics is not affected by the saddle-node bifurcation. In the opposite case, however, it can have dramatic consequences. This is related to the observation we made above, namely that a Hopf bifurcation can only occur for $Q_s y_s < \kappa$, while $Q_s y_s = \kappa$ is part of the conditions determining the saddle and the node. Even though these two conditions apply to two different fixed points, our numerical analysis strongly suggests that whenever there is a stationary state with $Q_s y_s = \kappa$ then the system does not have any stable limit cycle solutions. See Figure 4 for an example, where the limit cycle suddenly vanishes for $\kappa = \kappa_{\mathrm{sn}}$.

How does a limit cycle vanish as $\kappa \to \kappa_{\mathrm{sn}}$? First of all we observe that the limit cycle's amplitude does not change in this case, while its period $\tau$ diverges logarithmically, $\tau \sim |\log(\kappa - \kappa_{\mathrm{sn}})|$, see Fig. 4 inset. This behavior is characteristic for a saddle-loop or homoclinic bifurcation. In such a situation, a limit cycle collides with a saddle point, which results in a homoclinic orbit and the destruction of the limit cycle.

Let us finally remark that the disappearance of a limit cycle through a homoclinic bifurcation together with the appearance of a saddle-node suggests that the system is close to a Bogdanov-Takens bifurcation[21,22]. At such a bifurcation, which is of co-dimension 2, a saddle-node disappears. Furthermore, in the vicinity of the bifurcation point, a limit cycle appears via a Hopf bifurcation of the node. The cycle becomes unstable via a homoclinic bifurcation. However, we do not discuss this point any further, but rather turn to the case of a sarcomere.

## III. DYNAMICS OF A SARCOMERE

### A. Basic oscillation modes

A sarcomere consists of two identical half-sarcomeres that are rigidly coupled at their M-lines. The dynamic equations for a sarcomere are for one given by the force balance on each half-sarcomere. Explicitly, this leads to

$$f_{\mathrm{f},1} + 2(f_{\mathrm{e},1} + f_{\mathrm{m},1}) - (f_{\mathrm{e},2} + f_{\mathrm{m},2}) = 0 \qquad (13)$$
$$f_{\mathrm{f},2} + 2(f_{\mathrm{e},2} + f_{\mathrm{m},2}) - (f_{\mathrm{e},1} + f_{\mathrm{m},1}) = 0 \quad , \qquad (14)$$

where the indices 1 and 2 serve to distinguish between one and the other half-sarcomere. Applying the same approach as in Sect. II, we obtain dynamic equations for the linker extensions $y_i$, $i = 1, 2$, of the two half-sarcomeres:

$$\begin{pmatrix} g'(y_1) + 2\alpha_1 & -\alpha_2 \\ -\alpha_1 & g'(y_2) + 2\alpha_2 \end{pmatrix} \begin{pmatrix} \dot{y}_1 \\ \dot{y}_2 \end{pmatrix} = \begin{pmatrix} 2\beta_1 - \beta_2 \\ \beta_1 - 2\beta_2 \end{pmatrix} \qquad (15)$$

with $\alpha_i = \gamma\kappa Q_i(\varphi - \zeta\left[2g(y_i) + g(y_j)\right]/3)/(Q_j y_j - \kappa)$ for $j = 1, 2; j \neq i$ and $\beta_i = g(y_i) \cdot (Q_i y_i - \kappa) - \dot{Q}_i \alpha_i/(\gamma Q_i)$. The dynamic equations for the respective attachment probabilities $Q_i$, $i = 1, 2$, remain unchanged,

$$\dot{Q}_i = (1 - Q_i) - Q_i \cdot \omega(y_i) \quad . \qquad (16)$$

As we have seen in the previous section, half-sarcomeres can oscillate spontaneously and so one might - rightfully - expect sarcomeres to oscillate spontaneously, too. However, there is no parameter adjusting the coupling strength between the two halves of a sarcomere. Consequently, its dynamics cannot easily be understood on the ground of the results obtained for half-sarcomeres. Rather it has to be viewed as one self-sustained oscillator with proper oscillation modes. They do not result from synchronization of two independent oscillators.

After these words of caution, let us now discuss the dynamics of the sarcomere. For the analysis of the sarcomere dynamics, we will consider $\zeta$ as the control parameter. In this way, the results can readily be applied to the self-organized swimmer presented in Ref.[16]. A second reason for using $\zeta$ as control parameter will be given below. As above, changes of other parameter values can produce similar effects.

Figure 5 summarizes the bifurcation scenario upon variation of $\zeta$. The system's stationary states are given by those of the half-sarcomeres, because for the stationary states the sarcomere equations do decouple correspondingly. We will only consider the physically relevant situation when both half-sarcomeres are in the state determined by Eqs. (6). For sufficiently large values of $\zeta$, this state is stable. For decreasing values of $\zeta$, the system encounters a supercritical Hopf bifurcation at $\zeta = \zeta_1$ and starts to oscillate. In this state $\mathbf{O}_1$, both half-sarcomeres oscillate in the same way, but the oscillations are phase-shifted by half a period, $\varphi = \tau/2$[16]. $\mathbf{O}_1$ is invariant under simultaneous exchange of the half-sarcomeres and a time-shift of $\tau/2$. For $\zeta = \zeta_2$, there is a second Hopf bifurcation leading to a limit cycle $\mathbf{O}_2$. In this state, which is always unstable, the two half-sarcomeres oscillate synchronously, $\varphi = 0$. The critical values of the two oscillatory instabilities are given by

$$\zeta_1 = \frac{3\Lambda Q_s \left[1 + \omega(y_s) - y_s \omega'(y_s)\right]}{g'(y_s)(\kappa - Q_s y_s)[1 + \omega(y_s)]} - \frac{3(\kappa - Q_s y_s)}{\kappa [1 + \omega(y_s)]} \quad (17)$$
$$\zeta_2 = \zeta_1/3. \quad (18)$$

For $\zeta = \zeta_c$, the out-of-phase mode $\mathbf{O}_1$ loses stability via a pitchfork bifurcation. The frequency and the amplitude of the emergent limit cycle $\mathbf{Sw}$[23] is essentially the same as for $\mathbf{O}_1$ and also the form of the half-sarcomere oscillations remain quite similar. In contrast, the phase shift $\varphi$ does change. That is, $y_1(t) \approx y_2(t \pm \varphi)$ and $Q_1(t) \approx Q_2(t \pm \varphi)$ with $\varphi \neq \tau/2$. The slight differences between the oscillations of the two half-sarcomeres represent a spontaneous breaking of the $\mathcal{Z}_2$ symmetry of the dynamic equations. As a consequence two mutually symmetric solutions emerge at the bifurcation point. As the value of $\zeta$ is further decreased, the phase shift $\varphi$ further departs from $\tau/2$ but does not vanish in the limit $\zeta \to 0$. This is in accordance with the global instability of states with $\varphi = 0$. Let us note, that only for variations of $\zeta$ did we find two subsequent bifurcations. Changing $\gamma$ or $\kappa$ with all other parameters fixed, can either lead to a destabilization of the stationary state or of the oscillatory mode $\mathbf{O}_1$, but not of both in succession, further justifying our choice of the control parameter.

Similar to the canard explosion for a half-sarcomere, the limit cycles of a sarcomere can be strongly distorted for small changes of the control parameter. In this case, the dynamics on the limit cycle switches between small amplitude oscillations and large relaxation cycles. This is an example of mixed mode oscillations which are intimately linked to the canard phenomenon[24,25] and can indeed be interpreted as a generalization of the canard phenomenon[26]. Here, we do not investigate this phenomenon further.

### B. Gluing of cycles

We had seen above that half-sarcomeres can present homoclinic behavior. This remains true for the sarcomere, albeit in a different form. As the sarcomere's system parameters are changed so that it approaches a saddle-loop bifurcation, there are two distinct limit cycles that are about to be destroyed simultaneously. This is a consequence of the $\mathcal{Z}_2$ symmetry of the dynamic equations that has been broken spontaneously. However, instead of turning into homoclinic orbits after the bifurcation point, the two limit cycles fuse and form one common limit cycle. This phenomenon is called a gluing bifurcation[27,28].

In Figure 6A, we present the two limit cycles prior to a gluing bifurcation at $\kappa = \kappa_g$. The projection onto the $(Q_1, Q_2)$-plane, clearly reveals their symmetry. After the bifurcation, $\kappa \lesssim \kappa_g$, the two limit cycles have fused and look similar to the union of the two distinct limit cycles before the bifurcation, see Fig. 6B. That is, the $\mathcal{Z}_2$ symmetry is spontaneously restored! The scaling behaviour of the temporal periods at the bifurcation is again logarithmic from both sides, like in the half-sarcomere case, see Fig. 6C.

Gluing bifurcations are of co-dimension 2, so the variation of a single parameter, in our case $\kappa$, is generally insufficient to hit the bifurcation point. Inspection of the dynamics in the full phase space $(Q_1, Q_2, y_1, y_2)$ shows indeed that the system's behavior only resembles a gluing bifurcation. Rather, the phenomenon shown in Fig. 6 reflects ghost dynamics in the vicinity of a gluing bifurcation.

Note, that beside the gluing bifurcation, the $\mathcal{Z}_2$ symmetry of the dynamics can also be restored via the transition from state $\mathbf{Sw}$ to $\mathbf{O}_1$, which is shown in Fig. 5A.

### C. Chaotic behaviour

While the sarcomere states that we discussed above all have a counterpart in the dynamics of half-sarcomeres, there are states that are genuinely linked with the sarcomere. In fact, the limit cycle $\mathbf{Sw}$ can undergo further Hopf bifurcations, a phenomenon that is absent for half-sarcomeres. The second Hopf bifurcation generates a state $\mathbf{Sw}'$ with two incommensurate frequencies, while $\mathbf{Sw}$ becomes unstable, see Fig. 7A and B. Further bifurcations can occur. As a consequence, the state $\mathbf{Sw}'$ can either vanish through a backward Hopf bifurcation or acquire another frequency through another forward Hopf bifurcation. A still other possibility is the appearance of apparently chaotic solutions, see Fig. 7C. The system thus displays the classical Ruelle-Takens route to chaos via three subsequent Hopf bifurcations. After the third bifurcation even infinitesimal perturbations convert the periodic into chaotic motion. The chaotic regime extend only over a finite region in parameter space and ends at some finite value of $\zeta$.

## IV. CONCLUSIONS

In this work, we have studied a model for the dynamics of muscle sarcomeres. Our analysis shows that load-dependent detachment rates of myosin motors in conjunction with the elastic components present in sarcomeres leads to rich dynamic behavior of half-sarcomeres, including Hopf bifurcations, canards, and homoclinic bifurcations. Some of this behavior is rather common for relaxation oscillators and has been found also in other cytoskeletal processes involving force-dependent detachment rates. In the work by Enculescu et al[29], the dynamics of polymerizing actin filaments at a surface has been studied. There filaments can attach to the surface and detach with a load-dependent rate. For the analysis of that model, a similar mean-field approximation as used in the present work and as originally proposed by Grill et al[12] had been applied. The canard explosion as well as the homoclinic bifurcation should however be independent of this approximation and originate really from the detachment rate's load dependence.

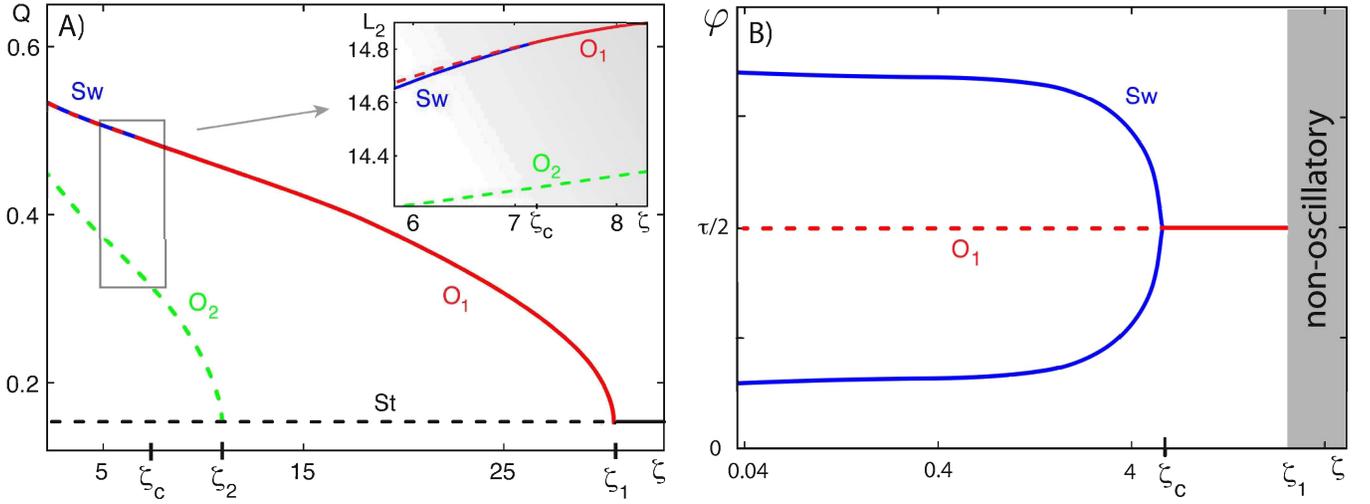

FIG. 5. Limit cycles of sarcomeres. A) Bifurcation diagram of the sarcomere. B) Relative phase $\varphi$ between the two half-sarcomeres for the limit cycles. Stable states are represented by continuous, unstable states by dashed lines. For $\zeta > \zeta_1$ the system has a stable stationary state **St**. For $\zeta = \zeta_1$, **St** loses stability and the oscillatory state **O$_1$** with $\varphi = \tau/2$ emerges. For $\zeta = \zeta_c$, **O$_1$** loses stability and the limit cycle **Sw** with $\varphi \neq 0$ emerges. For $\zeta = \zeta_2$ an unstable mode **O$_2$** with $\varphi = 0$ bifurcates from **St**. Inset: Detail of the bifurcation diagram around $\zeta = \zeta_c$ in terms of $L_2$ with $L_2 = \tau^{-1} \int_0^\tau \left(Q_1^2 + Q_2^2 + y_1^2 + y_2^2\right) dt$. Parameter values are $\gamma = 0.49$, $\Lambda = -9.9$, $M = 10$, and $\kappa = 0.16$. Other parameter values are as in Fig. 2.

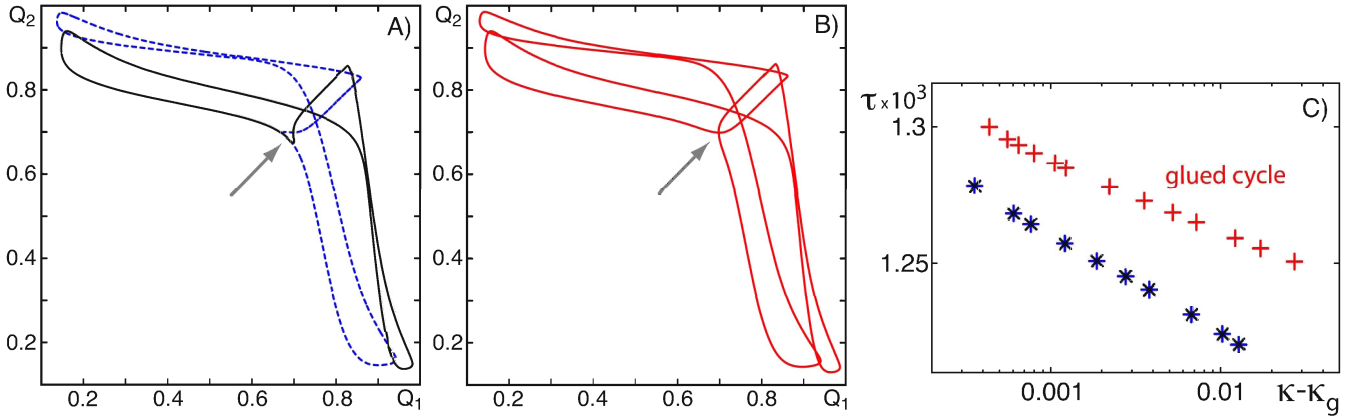

FIG. 6. Gluing bifurcation in a sarcomere. (A,B) Projection of the system's trajectories onto the $(Q_1, Q_2)$-plane. Arrows indicate where the cycles glue together. (A) The two mutually symmetric limit cycles for $\kappa = 0.64 \gtrsim \kappa_g$, (B) the limit cycle for $\kappa = 0.63 \lesssim \kappa_g$. (C) Period $\tau$ of the limit cycles before (blue) and after (red) gluing. Parameter values are as in Fig. 2 except for $\zeta = 33$.

The sarcomere, which can be viewed as two coupled half-sarcomeres, largely inherits the dynamic behavior of the half-sarcomere. However, due to the symmetry of the system, the homoclinic bifurcation turns into a gluing bifurcation and two limit cycles bifurcate from the stationary state: one with the two half-sarcomeres oscillating with a phase shift of $\tau/2$, the other with no phase shift. While the latter is always unstable, the former is stable until it looses stability through a pitchfork bifurcation, which leads to a change in the phase shift. This bifurcation leads to traveling contraction-relaxation waves along myofibrils. Our understanding of this bifurcation is currently limited. There is no coupling parameter that could be varied between the two half-sarcomeres. Consequently, this instability cannot be understood by a perturbation calculation.

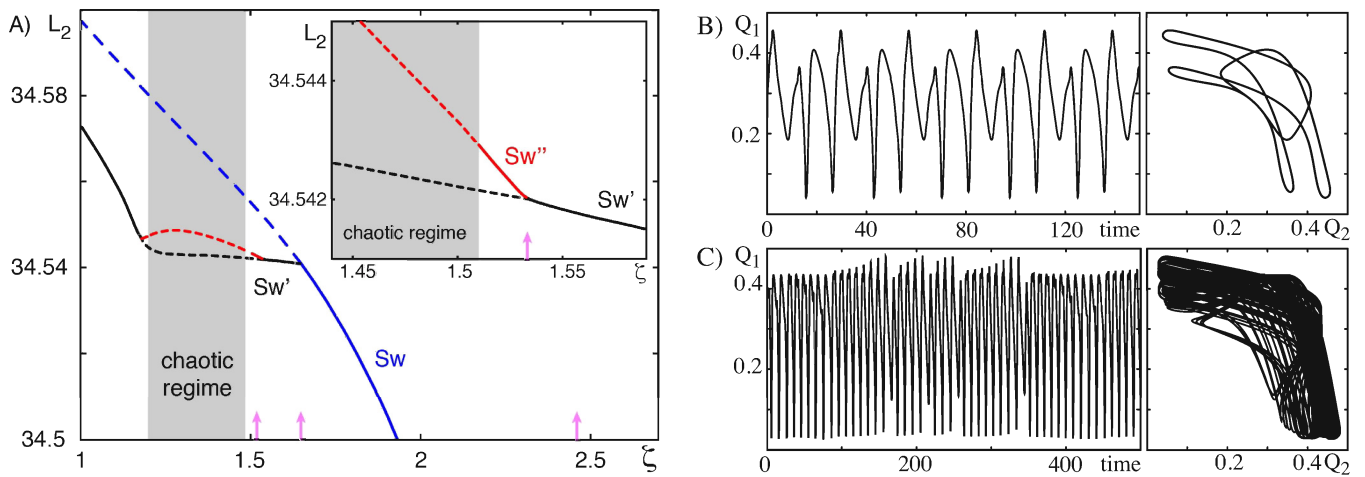

FIG. 7. Chaotic behaviour of sarcomeres. A) Bifurcation scenario according to the Ruelle-Takens route to chaos. Solid lines and dashed lines correspond to stable and unstable states, respectively. Primary and secondary Hopf bifurcations are indicated by arrows. The limit cycle **Sw** becomes unstable and the oscillatory solution **Sw'** emerges. **Sw'** undergoes another Hopf bifurcation towards the state **Sw''**. For decreasing $\zeta$ this state decays into an apparently strange attractor resulting in chaotic dynamics. Inset: Magnification of the transitions **Sw'** to **Sw''** and to chaotic dynamics. B) State **Sw'** as a function of time and in the $(Q_1,Q_2)$ plane for $\zeta = 1.6$. C) Example of a chaotic solution at $\zeta = 1.23$. Other parameters values are $\gamma = 0.05$, $\Lambda = -184.48$, $M = 7$, $\omega_0 = 0.16$, $\lambda = 1.61$, and $\kappa = 5.51$.

The bifurcation resembles the behavior of two pulse-coupled integrate-and-fire neurons[30]. In that system phase-shifts different from $\tau/2$ are generated if the coupling between the two neurons occurs via sufficiently short-lived pulses. In our system the analog of a pulse would be the force exerted by one half-sarcomere onto the other during the rapid relaxation. A formal mapping of our system onto the neuron case does not seem to be possible though.

A number of different proposals has been made to explain the spontaneous oscillations of muscle sarcomeres. They can be divided into two classes, those with a delayed force response and those with an anomalous force-velocity relation[31]. The mechanism we studied here belongs to the second class and is the only one we know of so far, which is capable of reproducing the relaxation waves along myofibrils. How can different mechanisms be distinguished experimentally? As we have seen above, the dynamics of our model is quite rich. Due to the canard explosion, it might be extremely hard to experimentally verify the Hopf bifurcation. For a myofibril, that is a chain of sarcomeres, the homoclinic bifurcations imply extremely slow propagation velocities of the relaxation waves. Stochastic effects are likely to mask this phenomenon. Also to distinguish chaotic behavior from stochastic dynamics is challenging. For these reasons, a quantitative comparison between our mechanism and experimental results requires a stochastic analysis. Still the different dynamic regimes found in our analysis of the mean-field model suggest a rich behaviour of sarcomeres that should have observable effects in the real system.

### ACKNOWLEDGMENTS

We would like to thank E. Nicola, A. Pikovsky, and M. Wechselberger for discussions.

[23] We denote the two states by **Sw**, because in the self-organized swimmer, this mode corresponds to swimming[16].
[2] J. W. Pringle, "The croonian lecture, 1977. Stretch activation of muscle: Function and mechanism." Proc R Soc Lond B, **201**, 107 (1978).
[3] N. Okamura and S. Ishiwata, "Spontaneous oscillatory contraction of sarcomeres in skeletal myofibrils." J Muscle Res Cell Motil, **9**, 111 (1988).
[4] S. Ishiwata, N. Okamura, H. Shimizu, T. Anazawa, and K. Yasuda, "Spontaneous oscillatory contraction (spoc) of sarcomeres in skeletal muscle." Adv Biophys, **27**, 227 (1991).
[5] C. J. Brokaw, "Molecular mechanism for oscillation in flagella and muscle." Proc Natl Acad Sci USA, **72**, 3102 (1975).
[6] F. Jülicher and J. Prost, "Cooperative molecular motors." Phys Rev Lett, **75**, 2618 (1995).
[7] F. Jülicher and J. Prost, "Spontaneous oscillations of collective molecular motors," Phys Rev Lett, **78**, 4510 (1997).
[8] D. Riveline, A. Ott, F. Jülicher, D. A. Winkelmann, O. Cardoso, J. Lacapere, S. Magnusdottir, J. Viovy, L. Gorre-Talini, and J. Prost, "Acting on actin: the electric motility assay," Eur Biophys J, **27**, 403 (1998).
[9] P.-Y. Plaçais, M. Balland, T. Guérin, J.-F. Joanny, and P. Martin, "Spontaneous oscillations of a minimal actomyosin system under elastic loading," Phys Rev Lett, **103**, 158102 (2009).
[10] T. Guérin, J. Prost, and J. F. Joanny, "Dynamic instabilities in assemblies of molecular motors with finite stiffness," Phys Rev Lett, **104**, 248102 (2010).
[11] S. Camalet and F. Jülicher, "Generic aspects of axonemal beating," New J Phys, **2**, 24.1 (2000).

[12] S. W. Grill, K. Kruse, and F. Jülicher, "Theory of mitotic spindle oscillations." Phys Rev Lett, **94**, 108104 (2005).

[13] O. Campas and P. Sens, "Chromosome oscillations in mitosis," Phys Rev Lett, **97**, 128102 (2006).

[14] S. K. Vogel, N. Pavin, N. Maghelli, F. Jülicher, and I. M. Tolic-Norrelykke, "Self-organization of dynein motors generates meiotic nuclear oscillations." PLoS Biology, **7**, 918 (2009).

[15] S. Günther and K. Kruse, "Spontaneous waves in muscle fibres," New J Phys, **9**, 417 (2007).

[16] S. Günther and K. Kruse, "A simple self-organized swimmer driven by molecular motors," Europhys Lett, **84**, 68002 (2008).

[17] E. Doedel, "Auto - software for continuation and bifurcation problems in ordinary differential equations," (2009-06-28).

[18] B. Ermentrout, *Simulating, analyzing, and animating dynamical systems* (Siam, 2002).

[19] E. Benoit, J. L. Callot, F. Diener, and M. Diener, "Chasse au canard," Collect Math, **32**, 37 (1981).

[20] M. Krupa and P. Szmolyan, "Relaxation oscillation and canard explosion," J Differ Equations, **174**, 312 (2001).

[21] F. Takens, "Forced oscillations and bifurcations," Comm Math Inst Rijksuniv Utrecht, **2**, 1 (1974).

[22] R. Bogdanov, "Bifurcations of a limit cycle for a family of vector fields on the plane," Selecta Math Soviet, **1**, 373 (1981).

[23] We denote the two states by **Sw**, because in the self-organized swimmer, this mode corresponds to swimming[16].

[24] A. Milik, P. Szmolyan, H. Loffelmann, and E. Groller, "Geometry of mixed-mode oscillations in the 3-d autocatalator," Int J Bifurcat Chaos, **8**, 505 (1998).

[25] M. Wechselberger, "Existence and bifurcation of canards in $\mathbb{R}^3$ in the case of a folded node," SIAM J Appl Dyn Syst, **4**, 101 (2005).

[26] M. Brons, M. Krupa, and M. Wechselberger, "Mixed mode oscillations due to the generalized canard phenomenon," Fields Inst Commun, **49** (2006).

[27] P. Coullet, J. M. Gambaudo, and C. Tresser, "A new bifurcation of codimension 2 - the gluing of cycles," C R Acad Sci I, **299**, 253 (1984).

[28] J. M. Gambaudo, P. Glendinning, and C. Tresser, "The gluing bifurcation: I. symbolic dynamics of the closed curves," Nonlinearity, **1**, 203 (1988).

[29] M. Enculescu, A. Gholami, and M. Falcke, "Dynamic regimes and bifurcations in a model of actin-based motility," Phys Rev E, **78**, 031915 (2008).

[30] L. F. Abbott and C. van Vreeswijk, "Asynchronous states in networks of pulse-coupled oscillators." Phys Rev E, **48**, 1483 (1993).

[31] A. Vilfan and E. Frey, "Oscillations in molecular motor assemblies," Journal of Physics: Condensed Matter, **17** (2005).